\documentclass[aps,prl,showpacs,twocolumn,floats]{revtex4}
\usepackage{amssymb}

\usepackage{dcolumn}
\usepackage{bm}

\begin{document}

\title{Origin of Spin Hall Effect (Reply to Comment)}

\author{E. M. Chudnovsky}
\affiliation{\mbox{Department of Physics and Astronomy,} \\
\mbox{Lehman College, City University of New York,} \\ \mbox{250
Bedford Park Boulevard West, Bronx, New York 10468-1589, U.S.A.}}
 \maketitle

Theory of spin Hall effect developed in my Physical Review Letter
\cite{Chudnovsky} is based upon non-relativistic one-particle
Hamiltonian with spin-orbit interaction to $1/c^2$:
\begin{equation}\label{ham1}
{\cal{H}} = \frac{{\bf p}^2}{2m} + U({\bf r}) +
\frac{\hbar}{4m^2c^2}{\bm \sigma}\cdot\left({\bm \nabla} U \times
{\bf p}\right)\,.
\end{equation}
Here $U({\bf r})$ is the electrostatic crystal potential felt by
charge carriers. With an accuracy to $1/c^2$ Hamiltonian
(\ref{ham1}) is mathematically equivalent to
\begin{equation}\label{ham2}
{\cal{H}} = \frac{1}{2m}\left({\bf p} - \frac{e}{c}{\bf
A}_{\sigma}\right)^2 + U({\bf r})
\end{equation}
where
\begin{equation}\label{A}
 {\bf A}_{\sigma} \equiv -\frac{\hbar}{4emc}\left({\bm
\sigma}\times {\bm \nabla}U\right)\,.
\end{equation}
Consequently, the orbital motion of electrons is affected by the
fictitious spin-dependent magnetic field:
\begin{equation}\label{B}
{\bf B}_{\sigma} = {\bm \nabla} \times{\bf A}_{\sigma} = -
\frac{\hbar}{4emc}\, \left[{\bm \nabla} \times \left({\bm
\sigma}\times {\bm \nabla}U\right)\right]\,.
\end{equation}
This fictitious field produces the same effect on the orbital
motion of electrons as the real magnetic field does in the
conventional Hall effect, but with the Hall currents having
opposite directions for electrons with opposite spin
polarizations.

Equations (\ref{ham1}) - (\ref{B}) provide conceptual framework
needed to understand the spin Hall effect. According to Eq.\
(\ref{B}), for the intrinsic spin Hall effect to exist, one needs
to satisfy two requirements. First, charge carriers should be able
to traverse the sample before scattering reverses their spin. This
can be easily achieved in a small sample at low temperature.
Second, ${\bm \nabla}_i{\bm \nabla}_jU$ should have a non-zero
average over electron states. Since the homogeneous external
electric field does not contribute to this average, the effect
must be entirely due to the inhomogeneity of the crystal field.
For a cubic crystal one obtains
\begin{equation}\label{cubic}
\langle{\bm \nabla}_i{\bm \nabla}_jU\rangle = C\delta_{ij}
\end{equation}
due to the cubic symmetry alone, with $C$ being a constant. This
constant was estimated in my Letter from the Laplace equation:
\begin{equation}\label{Laplace}
C \equiv \frac{1}{3}\langle {\bm \nabla}^2 U({\bf r}) \rangle = -
\frac{4 \pi}{3} e \langle \rho({\bf r}) \rangle\,,
\end{equation}
where $\rho({\bf r})$ is the charge density that creates $U({\bf
r})$.

Kravchenko \cite{Kravchenko} argues that the right hand side of
Eq.\ (\ref{Laplace}) must be zero due to electric neutrality of
the solid. This argument is incorrect. It is contrary to the
conventional approach to solids in which electron states are
formed by the potential due to localized charges arranged in a
crystal lattice \cite{AM}. Such a potential was chosen in my
Letter. It is in line with the fact that spin-orbit interaction is
large when electron passes close to the localized charge. {\it
Electric neutrality, that is, the screening of the localized
charges by conduction electrons, occurs at greater distances}.
Taken literally, without relevance to the spatial scale, the
electric neutrality would prohibit existence of solids. Without
coupling of electrons to a localized (only partially screened)
positive central charge even individual neutral atoms would not
exist. In the same way, the screening of the localized charge by
conduction electrons at large distances is irrelevant in the
context of spin-orbit interaction in Eq.\ (\ref{ham1}) that leads
to the spin Hall effect.

In my Letter I chose $U({\bf r})$ created by a cubic lattice of
ions of charge $-Ze > 0$. For such a choice $\langle \rho \rangle
= -Zen_0 = -en$ where $n_0$ and $n=Zn_0$ are concentrations of
ions and conduction electrons, respectively. This gives
\begin{equation}\label{B-final}
\langle {\bf B}_{\sigma} \rangle =\frac{4\pi}{3}n \mu_B {\bm
\sigma}\,,
\end{equation}
with $\mu_B$ being the Bohr magneton. As is shown in the Letter
\cite{Chudnovsky}, this result provides the magnitude of spin Hall
conductivity that is in quantitative agreement with experiments in
cubic metals and semiconductors. Note that $\langle {\bf
B}_{\sigma} \rangle$ depends on the crystal symmetry and should
not be confused with the magnetic field created by polarized
electrons. For, e.g., a tetragonal crystal, the average
$\langle{\bm \nabla}_i{\bm \nabla}_jU\rangle$ should be of the
form $C\delta_{ij} + Dn_in_j$ (with ${\bf n}$ being a tetragonal
axis), which gives $\langle {\bf B}_{\sigma} \rangle$ of the form
$C'{\bm \sigma} + D'({\bf n}\cdot{\bm \sigma}) {\bf n}$.
Consequently, in a tetragonal crystal with polarized electrons,
the electric current due to the external electric field should be
\begin{equation}
{\bf j} = \sigma_c {\bf E} + \sigma_{s1}[{\bm \xi} \times {\bf E}]
+ \sigma_{s2}({\bm \xi}\cdot{\bf n})[{\bf n} \times {\bf E}]\,,
\end{equation}
where $\sigma_c$ is charge conductivity, and $\sigma_{s1},
\sigma_{s2}$ are two spin Hall conductivities; ${\bm \xi}$ being
the polarization ($0 < \xi < 1$) of the electrons. Similar
expressions can be obtained for crystal lattices of arbitrary
symmetry. This prediction of the theory on how the spin Hall
current depends on the symmetry and orientation of the crystal can
be tested in experiment.

The intrinsic spin Hall effect described above, and in my PRL
\cite{Chudnovsky}, is a crystal counterpart of spin-dependent
(Mott) scattering by individual (neutral) atoms \cite{Mott}.
Kravchenko's Comment fails to appreciate that both effects require
spatial inhomogeneity of the electric field. This requirement was
also ignored by researchers who employed the so-called Rashba
spin-orbit interaction. The latter corresponds to Eq.\
(\ref{ham1}) with ${\bm \nabla}U$ replaced by a constant. While
some argument in favor of such a replacement can be made for a
purely two-dimensional electron system placed in a strong
transverse electric field, it is certainly incorrect for a
three-dimensional system. As can be seen from the above formulas,
the replacement of ${\bm \nabla}U$ with a constant results in
${\bf B}_{\sigma} = 0$. This makes Rashba spin-orbit interaction
unsuitable for the description of the intrinsic spin Hall effect.

This work has been supported by the DOE grant No.
DE-FG02-93ER45487.


\begin{thebibliography}{10}
\bibitem{Chudnovsky}
E. M. Chudnovsky, Phys. Rev. Lett. {\bf 99}, 206601 (2007).
\bibitem{Kravchenko}
V. Ya. Kravchenko, preceeding Comment, Phys. Rev. Lett. {\bf 100},
199703 (2008).
\bibitem{AM}
See, e.g., N. W. Ashcroft and N. D. Mermin, {\it Solid State
Physics}, Chapter 8, (Holt, Rinehart, and Winston, New York,
1976).
\bibitem{Mott}
N. F. Mott, Proc. R. Soc. A {\bf 124}, 425 (1929).
\end{thebibliography}
\end{document}